\documentclass[%
superscriptaddress,
preprint,
 amsmath,amssymb,
 aps,
 pra,
floatfix,
citeautoscript
]{revtex4-2}

\usepackage{graphicx}% Include figure files
\usepackage{dcolumn}% Align table columns on decimal point
\usepackage{bm}% bold math
\usepackage{chemformula} % Formula subscripts using \ch{}
\usepackage[T1]{fontenc} % Use modern font encodings
\usepackage{amssymb}
\usepackage{bm}
\usepackage{amsmath}
\usepackage{accents}
\usepackage{booktabs}
\usepackage{dcolumn}
\usepackage{chemfig}
\begin{document}

%\preprint{APS/123-QED}

\title{Study of the Molecular Level Mechanism of Nanoscale Alternating Current Electrohydrodynamic Flow. }% Force line breaks with \\

\author{Sobin Alosious}
\affiliation{Australian Institute of Bioengineering and Nanotechnology, The University of Queensland, Brisbane, QLD 4072, Australia}
%\email{s.alosious@uq.edu.au}
\author{Fiach Antaw}
\affiliation{Australian Institute of Bioengineering and Nanotechnology, The University of Queensland, Brisbane, QLD 4072, Australia}
%\email{f.antaw@uq.edu.au}
\author{Matt Trau}
\affiliation{Australian Institute of Bioengineering and Nanotechnology, The University of Queensland, Brisbane, QLD 4072, Australia}
\affiliation{School of Chemistry and Molecular Biosciences, The University of Queensland, Brisbane, QLD 4072, Australia}
%\email{m.trau@uq.edu.au}
\author{Shern R. Tee}
\affiliation{School of Environment and Science, Griffith University, Nathan, QLD, 4111, Australia}
%\email{s.tee@uq.edu.au}
\author{Debra J. Searles}

\email{d.bernhardt@uq.edu.au}
\affiliation{Australian Institute of Bioengineering and Nanotechnology, The University of Queensland, Brisbane, QLD 4072, Australia}
\affiliation{School of Chemistry and Molecular Biosciences, The University of Queensland, Brisbane, QLD 4072, Australia}
\affiliation{ARC Centre of Excellence for Green Electrochemical Transformation of Carbon Dioxide, The University of Queensland, Brisbane, QLD 4072, Australia}



\begin{abstract}
This study investigates the molecular-level mechanism of Alternating Current Electrohydrodynamic (AC-EHD) flow in nanopores under high-frequency conditions, using molecular dynamics simulations. A gold-NaCl system with symmetric and asymmetric electrode configurations is used to analyze the flow patterns under high-frequency AC potentials. Our findings reveal localized heat generation near the electrode leading to a steep temperature gradient. An order parameter analysis indicates that the heat generation is due to the periodic change in the alignment of water molecules under AC potentials. At these high frequencies the influence of Na$^+$ and Cl$^-$ ions are negligible. The heat generation and temperature gradient are found to increase with the applied AC frequency. Three different electrode configurations were studied by varying the size and distance between the electrodes. A net directional flow develops in the asymmetric electrode structures. A possible mechanism for this is proposed by analyzing the flow patterns using velocity and temperature profiles, order parameters, streamline plots and mean square displacements. Different effects on the fluid were identified including those associated with temperature gradients, temperature-dependent fluid properties, and non-uniform electric fields. The asymmetric electrode structure created an imbalance in these effects and generated a net directional flow. These findings suggest the existence of a form of nanoscale AC-EHD flow that operates in a frequency regime above that of conventional electroosmotic and electrothermal mechanisms and that, unlike these mechanisms, occurs independently of ionic concentration. Thereby this work provides insights for optimizing AC-EHD flow in nanoscale systems where precise fluid manipulation is critical.

%\begin{description}
%\end{description}
\end{abstract}

\maketitle

%\tableofcontents

\section{INTRODUCTION}

Electrohydrodynamics (EHD) is an interdisciplinary science between electrical and fluid mechanics that studies the dynamics of electrically charged or polarizable liquids due to the action of electric fields. The domain of EHD has potential for substantial theoretical and practical application ranging from microfluidics and the enhancement of heat transfer to electrostatic precipitation \cite{jones1995electromechanics,melcher1969electrohydrodynamics,velev2006chip,bazant2004induced}. Interaction between charges or dipoles in a fluid medium either creates flow or amplifies flow patterns, thus enabling the alteration of fluid behavior without using mechanical components \cite{castellanos2014electrohydrodynamics}. Alternating Current Electrohydrodynamic (AC-EHD) flow represents one branch of EHD that applies AC electric fields to induce fluid motion. Applying an AC potential prevents the electrolysis and electrode fouling problems, which are typical in DC systems. This makes it much more suited for use in delicate biochemical processes or wherever chemical purity is a concern.
The AC-EHD method utilizes time-averaged forces, such as dielectrophoresis and electrothermal forces, which are quite different from those used in DC-EHD systems \cite{ramos1999ac}. The use
of AC-EHD is particularly notable in microscale applications where precise control over fluid mixing, heat transfer, and mass transport is essential.  These capabilities are necessary for the development of technologies like lab-on-a-chip devices, precision cooling systems, and micro-reactors \cite{ramos1998ac,nguyen2006fundamentals,silva20223d,ronkainen2010electrochemical}. The recent advancements in understanding the AC-EHD phenomena have led to have led to the design of complex systems that employ the advantages of their fluid dynamics for better performance.

Over the years, crucial technological advances have harnessed the potential of AC-EHD and electrokinetic forces for isolation and detection of exosomes \cite{wei2013detection,boriachek2018biological}, DNA trapping and analysis \cite{basuray2009shear,kumemura2011single}, protein biomarker analysis \cite{gong2010label,vaidyanathan2015multiplexed}, cancer cell isolation and analysis \cite{gupta2012apostream,dey2016electric}, and manipulation of cells and microorganisms \cite{kwon2018microfluidic,ren2018bioparticle}. These advances have enabled sensitive detection and precise manipulation of clinically relevant biomarkers from complex biological samples. Shiddiky et al. \cite{shiddiky2014molecular}  introduced a technique called AC-EHD-induced nano shearing to improve the detection of molecular biomarkers in complex biological samples. This method uses an alternating electric field to displace molecules that are nonspecifically bound to solid surfaces. The force generated by the electric field acts within nanometers of the electrode surface, shearing off loosely bound molecules and reducing nonspecific adsorption by up to five times. Additionally, this technique enhances the specific capture of target proteins, leading to a 1000-fold increase in the detection of HER2 protein from human serum, with sensitivity down to 1 fg mL$^{-1}$. This approach allows for controlled shear forces and fluid micro-mixing, improving sensor-target interactions and making it suitable for clinical applications and early disease diagnosis.

Vaidyanathan et al. \cite{vaidyanathan2014detecting} have developed a microfluidic platform by using AC-EHD induced nano shearing for capturing and detection of highly specific exosomes which is crucial for cancer biomarker analysis. This method utilizes AC-EHD-induced electrical forces near the electrode surface to enhance the capture specificity and reduce nonspecific adsorption. The device effectively isolated exosomes from breast cancer samples and showed that the detection sensitivity is three times better than hydrodynamic flow assays.  
Their platform using asymmetric pairs of microelectrodes is a promising tool for the rapid quantification of exosomes, diagnosis of cancer and other biological applications.

Researchers have employed various experimental and analytical methodologies to understand the mechanism behind AC-EHD flow, thereby optimizing the conditions for its effective implementation. 
Brown et al. \cite{brown2000pumping} consider asymmetric electrodes where flow is induced by a gradient in the potential parallel to their surface. The voltages required to observe this are relatively small, and the dependence of the flow on the potential and frequency is studied in this work. A model is proposed that is able to qualitatively predict the observed fluid velocities, and factors that need to be considered to improve the model are described. 

Ramos et al. \cite{ramos2003pumping} conducted a theoretical analysis of electro-osmotic pumping which is driven by a kilohertz-scale AC electric potential applied across pairs of asymmetric microelectrodes. A net fluid flow is generated due to the interaction between the oscillating electric field and the induced charges in the diffused double layer of the electrode. They numerically solve the electrical equation using the charge simulation method, and the computational efficiency in the calculation is enhanced by considering the periodicity of the system. Optimal non-dimensional parameters are identified for enhancing the pumping velocity and the dependence of the fluid flow on voltage and frequency is determined. Their results are well aligned with the experimental data reported by Brown et al. \cite{brown2000pumping}.

González et al. \cite{gonzalez2006electrothermal} studied electrothermal motion in aqueous solutions subject to both megahertz-scale frequency alternating electric fields and temperature gradients produced by illumination of the system. They have developed solutions for fluid flow using two types of microelectrode structures. The first structure is two coplanar electrodes with an AC potential difference which generates two-dimensional rolls, and the second structure is four planar coplanar electrodes with  a four-phase AC signal, creating a rotating liquid whirl. 
The theoretical predictions are confirmed by the experiments under strong illumination. Their findings suggest that such electrothermal flows could enhance mixing in microsystems where the flow velocity is independent of the solution conductivity and the efficiency can be improved by tuning the AC frequency.

Different mechanisms have been reported to explain AC-EHD flow across various frequency ranges. Among these, two distinct mechanisms are particularly prominent in the literature. The first mechanism involves applying an alternating potential difference across asymmetric electrodes, creating a nonuniform electric field. The normal component of the field will induce a charge in the diffuse double-layer and the tangential component will exert a force on this induced charge. 
Since the sign of the charge reverses with the field, the force has a nonzero time average which results in an electro-osmotic slip velocity with both oscillatory and time-averaged components. The time-averaged velocity predominates in the the hundreds to the kilohertz range frequencies causing an observed fluid flow.  This setup with coplanar asymmetric electrodes produces a localized flow that generates a global flow in the direction of broken symmetry.\cite{ramos2003pumping,ajdari1996generation,ramos1999ac,green2002fluid}

The second mechanism is electrothermally induced fluid flow, in which the electric fields interact with temperature gradients within an aqueous solution, causing fluid motion.  These temperature gradients can be generated by different external sources, such as strong illumination, or through Joule heating from the applied electric field. The gradients will create an inhomogeneity in the system due to the variation of electrical conductivity and permittivity of fluid. When an alternating current electric field is applied, the forces induced lead to fluid motion. Depending on the electrode configuration, this mechanism can produce different flow patterns: two coplanar electrodes with an AC potential difference generate two-dimensional rolls, while four coplanar electrodes with a four-phase AC signal create a rotating whirl. This electrothermal motion is useful in microsystems for mixing analytes and liquids, which offers advantages such as efficiency and independence of flow velocity from solution conductivity \cite{gonzalez2006electrothermal,green2000electric,green2001electrothermally}. Even though various theories exist in the literature explaining the underlying mechanism of AC-EHD flow, there remains a need for more accurate molecular-level explanations, especially at higher frequencies and nanoscale dimensions. Further research is essential to fully understand and optimize the flow for various applications.  Recent numerical studies have further advanced the understanding of AC-driven electrohydrodynamic phenomena at micro- and mesoscopic scales, highlighting the roles of AC frequency, interfacial polarization, and electric-field nonuniformity in governing flow and particle dynamics. In particular, Tao et al. demonstrated frequency-dependent electrohydrodynamic interactions and Maxwell-stress-mediated transport using fully coupled multiphysics
simulations~\cite{tao2025many,tao2025alternating}. These studies provide important context for AC-EHD behavior at larger length scales and motivate the need for molecular-level investigation under extreme nanoscale
confinement.

Experimental studies have demonstrated that  AC-EHD excitation can generate localized near-surface fluid motion over asymmetric electrodes, often referred to as ``nanoshearing'', even though the overall device dimensions remain at the microfluidic scale. In particular, Shiddiky et al. ~\cite{shiddiky2014molecular} and Vaidyanathan
et al.~\cite{vaidyanathan2014detecting} reported AC-EHD-induced near-electrode flow confined to within a few nanometers of the electrode surface under kHz--MHz excitation. While these experiments confirm the physical
existence of AC-EHD-driven near-surface flow, controlled measurements under extreme nanoscale confinement and GHz-frequency excitation are not currently available, motivating molecular-level investigation.

As systems reduce in size to the nanoscale, the mechanisms are likely to differ from those proposed previously. In this paper, we investigate the molecular-level mechanism of nanoscale AC-EHD flow with a gold-NaCl aqueous system using molecular dynamics simulations. Molecular dynamics is used to capture molecular-scale and interfacial mechanisms that govern electrohydrodynamic flow under extreme nanoscale confinement, which are not explicitly resolved in mesoscopic approaches.
An AC potential with a range of very high-frequency range was employedfrequencies is applied to the systems using the constant potential method to analyze (CPM) and the flow patterns. are analyzed. Various electrode arrangements wereare tested to determine how the optimum geometry for achievingcan be tuned to achieve maximum flow velocity. Additionally, a theoretical explanation for AC-EHD flow at the nanoscale wasis developed based on the observed results.
The following sections provide a summary of the modeling approach, the simulation methodology employed, a discussion of the obtained results, and, finally, some concluding remarks.

\section{METHODOLOGY}
\begin{figure}[h!]
\centering
\includegraphics[width=.9\textwidth]{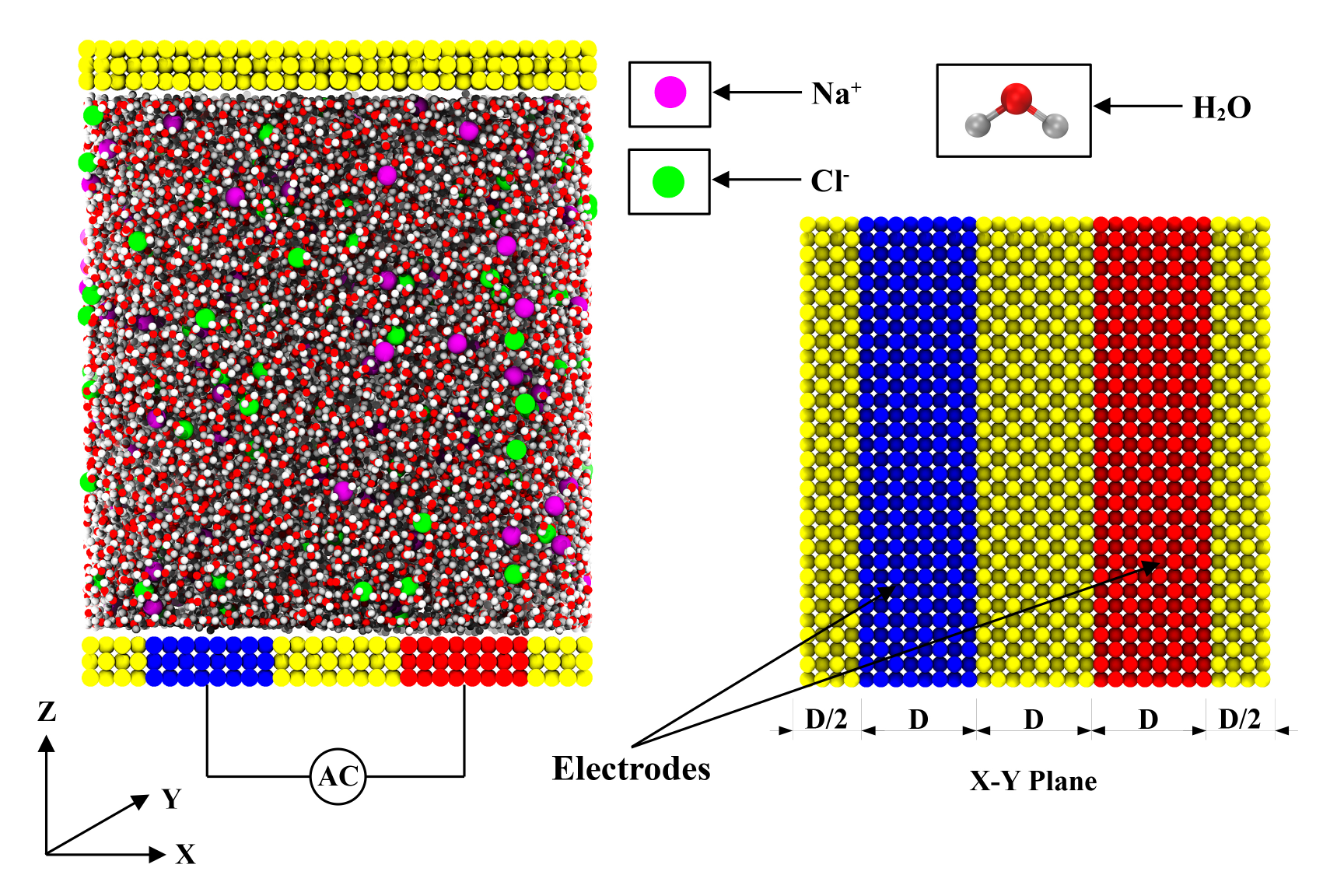}
\caption{Schematic diagram of a typical simulation setup.  }
\label{model}
\end{figure}

Classical molecular dynamics (MD) simulations were carried out to investigate the underlying mechanism of alternating current electrohydrodynamic flow. The simulation setup consists of an aqueous solution of NaCl confined between two gold nanoplates. The schematic depiction of one of the models (model 1) used in this study is shown in Fig. \ref{model}. The atoms in the top Au plate are flexible, with the Au atoms  tethered around their initial lattice positions using a weak harmonic spring potential. The bottom gold plate contains electrodes placed at a fixed distance with rigid atoms. The dimensions of the Au sheets are L$_x$ = L$_y$ = 5.5~nm and the height of the channel is  L$_z$ = 7~nm. The length of both electrodes is D = 1.375~nm and the gap between the electrodes is also D = 1.375~nm for the model shown. Other models are also investigated with different electrode sizes and gaps between the electrodes. Periodic boundary conditions were applied in the $x$ and $y$ directions, and confinement was in the $z$-direction

The water molecules were modeled using the extended simple
point charge (SPC/E) \cite{berendsen1987missing,wu2006flexible}  water model. The long-range electrostatic forces were calculated by using the
particle-particle-particle-mesh (PPPM)  \cite{hockney2021computer} solver with an accuracy of 1 $\times$ 10$^{-6}$, and the water molecules were kept rigid by using the SHAKE  \cite{ryckaert1977numerical} algorithm. To use the PPPM method, periodic boundary conditions in all directions are required. An Ewald summation technique with an extended volume ratio of 3.0 was used due to the confinement in the $z$ direction \cite{yeh1999ewald}. The non-Coulombic interactions between particles are of the Lennard-Jones form and the parameters for the cross-terms are derived using the Lorentz-Berthelot mixing rules. The force field parameters for Au(100) were taken from Wright et al. \cite{wright2013golp}. The non-polarizable NaCl with SPC/E water models were modeled using the parameters taken from Weerasinghe and Smith \cite{weerasinghe2003kirkwood}.  A velocity-Verlet algorithm was used to integrate the equations of motion for every particle, with a time step of 1 fs. All of the MD simulations were carried out using the LAMMPS (large-scale atomic/molecular massively parallel simulator) package \cite{plimpton1995fast}, while the CPM was carried out using the ELECTRODE \cite{siepmann1995influence} package. Visual MD \cite{humphrey1996vmd} is used for model rendering and visualization.

Initially, the system was equilibrated in a canonical (NVT) ensemble at a reference temperature of 300 K for a time period of 2.0 ns. In
addition, the system stability was checked by simulating the system in
a microcanonical (NVE) ensemble for another 2.0 ns. Once the system was properly equilibrated, a constant potential method was applied to the electrodes, and the top gold plate was thermostatted at 300 K for 2 ns. Finally, all the analyses were carried out during another 2 ns production simulation by maintaining the electrode potential and thermostat. A 4 V potential difference was applied to the electrodes using the constant potential method for both AC and DC. The AC frequency was varied from a few MHz to 100 GHz to understand the range of frequency required to make any impact on the system at this timescale. Since the time scale used in this molecular dynamics simulation was in the nanosecond range, at least 1 GHz frequency is required to make a complete cycle of AC potential for 1 ns time. The cumulative heat removed from the system is calculated using the energy removed from the thermostat.

\section{RESULTS AND DISCUSSIONS}
\begin{figure}[h!]
\centering
\includegraphics[width=.9\textwidth]{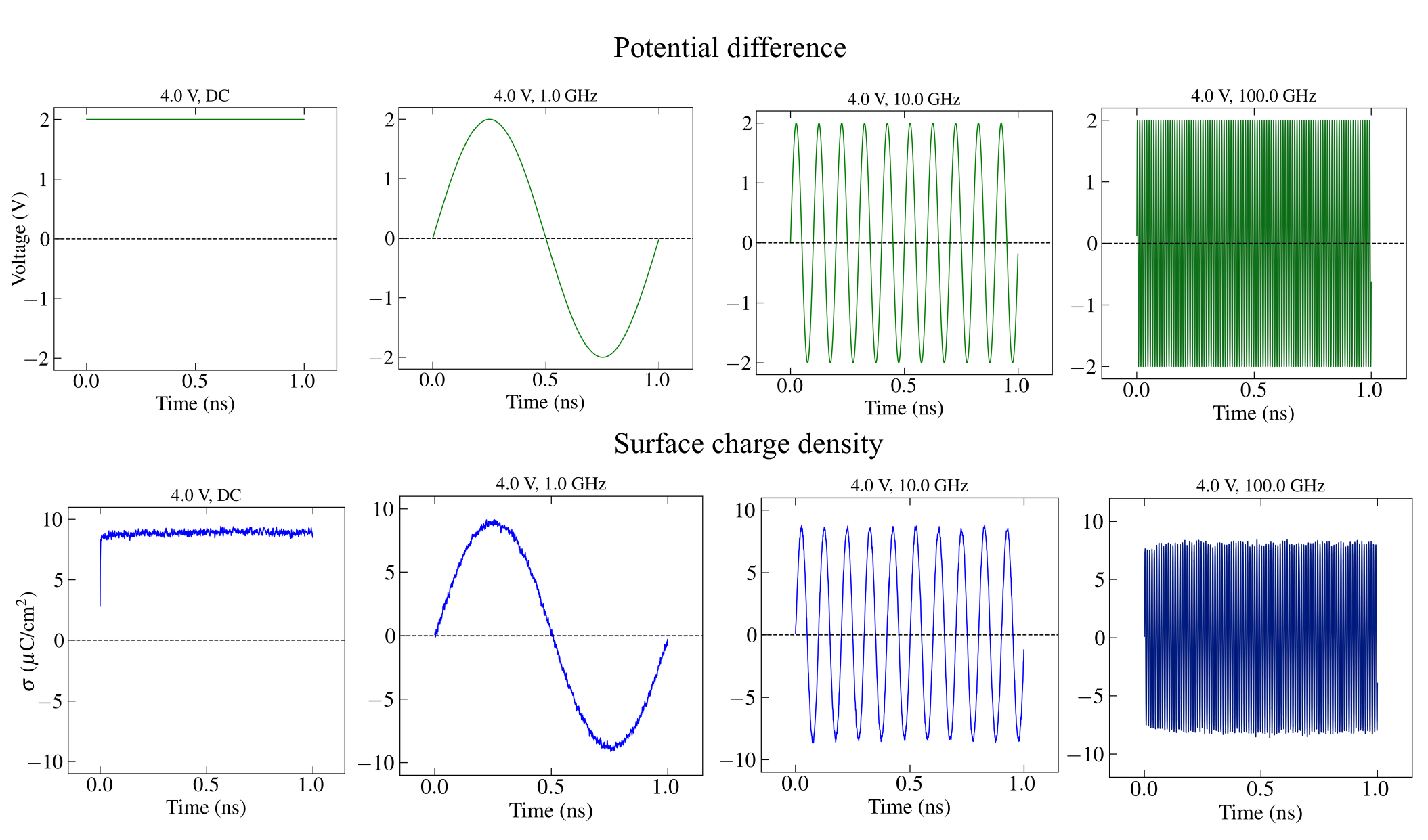}
\caption{The applied potential difference and the corresponding surface charge density as a function of time for 4.0 V DC, 1 GHz, 10 GHz, and 100 GHz AC voltages. }
\label{frequency}
\end{figure}

The surface charge density of the electrodes with different frequencies was investigated for frequencies from 1 MHz to 100 GHz. For the lower frequencies, the change in polarity during the time scale used in this study (nanoseconds) is not enough to have any observable influence on the system. Therefore we used very high frequencies -- up to 100 GHz -- to understand the molecular level mechanism of AC-EHD flow. Figure \ref{frequency} shows the applied voltage and induced surface charge as a function of simulation time for a few selected frequencies. For the case of DC voltage, surface charge density starts from zero and reaches the maximum value of 4.0 V and remains constant throughout the simulation time. For a frequency of 100 MHz, the period is 10 ns so the voltage remains small during the 1 ns simulation time. For frequencies lower than this, it will take a very large simulation time to obtain the peak value of the voltage (first peak of sine wave). A 250 MHz frequency range is required to reach the peak voltage in 1 ns. From Fig. \ref{frequency}  we can see that a complete cycle is obtained when the frequency is 1 GHz. Therefore to study the effect of frequency on the system using molecular simulations in a viable timeframe, we need a very high frequency. From Fig \ref{frequency}  it is also evident that the response time of induced charge is very fast, as the surface charge density nearly follows the voltage curve even for higher frequencies. 

\begin{figure}[h!]
\centering
\includegraphics[width=1.0\textwidth]{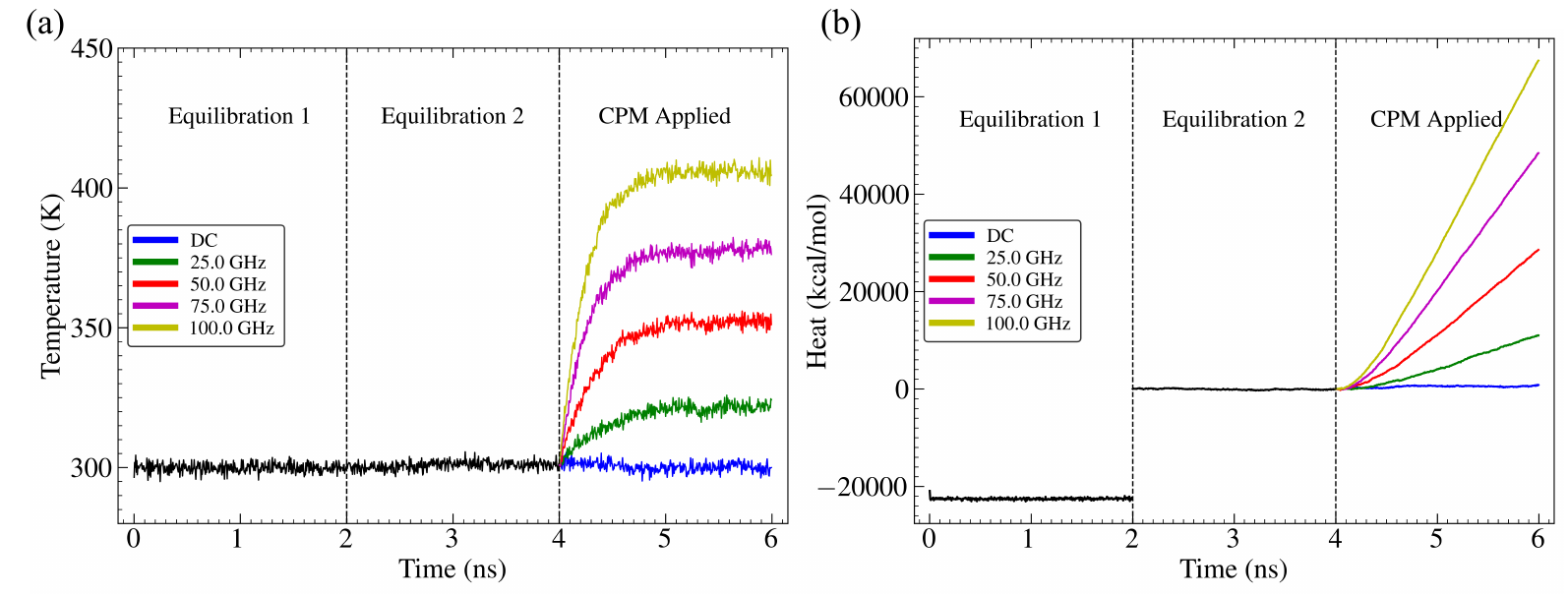}
\caption{(a) Temperature of the fluid as a function of simulation time for different frequencies. (b) Heat removed by the thermostat as a function of simulation time for different frequencies. }
\label{temp_frequency}
\end{figure}

As mentioned in the methodology section, the AC potential is applied after the equilibration of the system. Once the constant potential is applied, the induced charge on the electrode changes its polarity according to the frequency of the AC potential applied. While analyzing the temperature of the system it was observed that the system temperature is increasing with time and reaches a steady state value after a particular time. Figure \ref{temp_frequency} shows the temperature of the fluid and heat removed by the thermostat for DC voltage and different AC voltage frequencies. In the case of DC voltage, the temperature of the fluid remains the same throughout the simulation period. However, it is observed that when the frequency of the AC is approximately 1 GHz, the temperature of the fluid can be observed to increase and reach a steady state. The temperature rise is proportional to the frequency of the applied AC potential. The increase in fluid temperature indicates that heat is generated in the system while applying AC potential, as would be expected.  Figure \ref{temp_frequency}b shows the heat removed by the thermostat for different frequencies. The heat removed by the thermostat remains constant for DC potential indicating zero heat generation after some time. When the AC potential is applied, the heat removed by the thermostat increases over time, indicating continuous heat generation that also depends on the frequency.  We note that the observed heating is spatially localized near the electrodes, present even in low-flow configurations, and results in steady-state temperatures roughly proportional to the applied AC frequency (\ref{temp_frequency}(a)). Thus, while our molecular dynamics simulations should include most physically relevant energy dissipation mechanisms, including viscous dissipation due to fluid motion, the observed heating is directly attributable to the AC frequency imposed rather than resulting predominantly from viscous dissipation.

\begin{figure}[h!]
\centering
\includegraphics[width=1\textwidth]{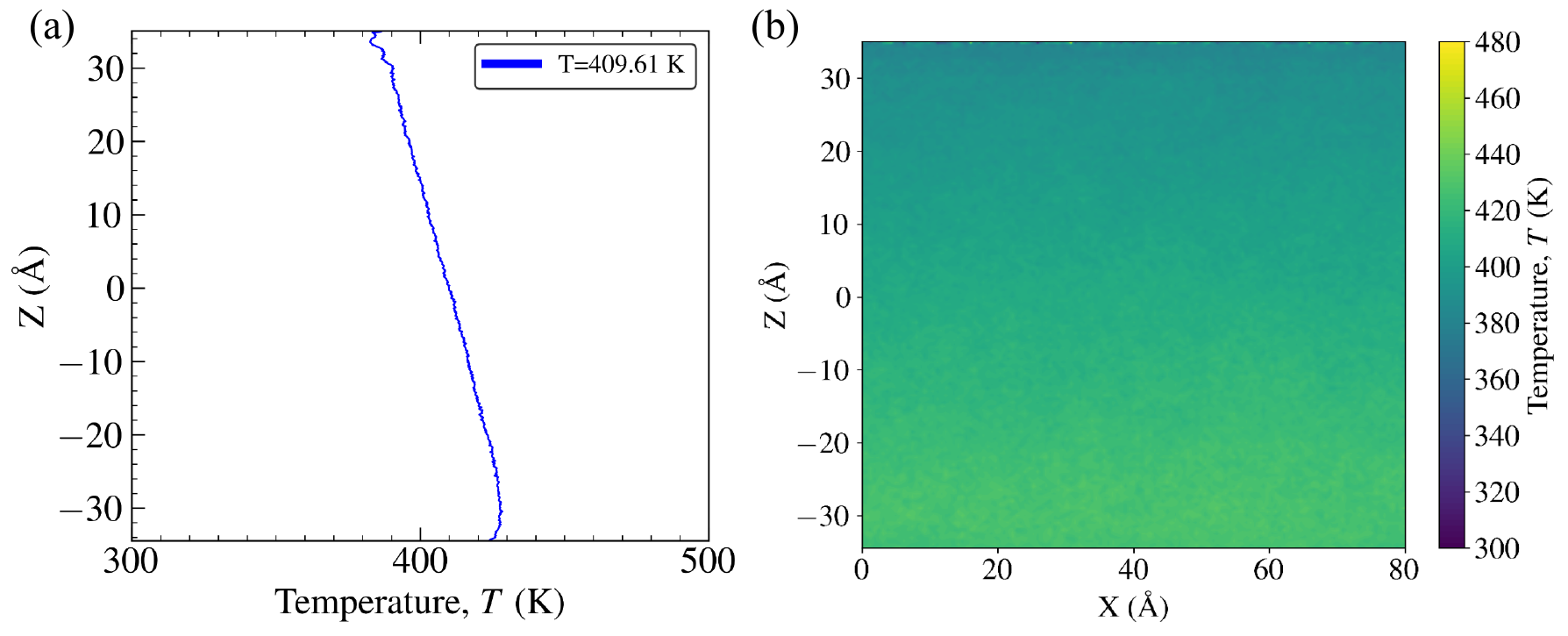}
\caption{(a) Temperature profile along the $z$ direction (top to bottom). The average temperature is 409.61 K. (b) Temperature heat map in the $x$-$z$ plane.  }
\label{temp_contour}
\end{figure}

To understand the pattern of heat generation in the system, the temperature profile and heat map of one of the systems are plotted in Fig \ref{temp_contour}. The temperature profile is determined by calculating local temperatures using the degrees of freedom correction method, which accounts for constraints in the molecular system\cite{sanderson2024local,sanderson5187317dofulator}. The temperature profile indicates that the temperature is maximum at the bottom plate, where the electrodes are placed, and gradually decreases towards the top plate. Therefore, it is understood that heat generation does not occur in the fluid as a whole; instead, heat is generated locally near the electrodes. In addition, the temperature contour plot shows that there are no heat generation spots along the electrode surface; instead, a uniform increase of temperature over the electrodes is visible. This could change if the electrodes increase in size.
\

\begin{figure}[h!]
\centering
\includegraphics[width=1\textwidth]{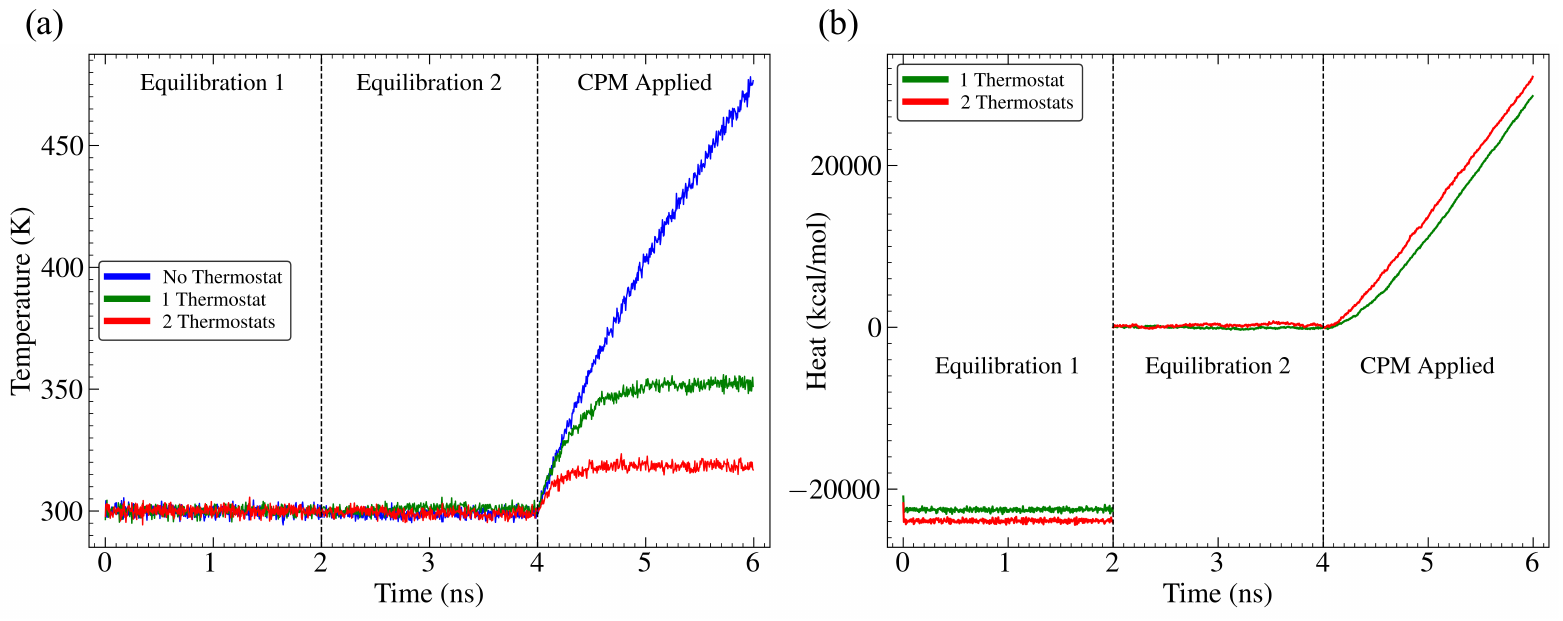}
\caption{(a) Temperature of the fluid as a function of simulation time for different thermostat settings. (b) The heat removed by the thermostat as a function of simulation time for different thermostat settings. }
\label{temp_thermo}
\end{figure}

 As mentioned in the methodology section, for the simulations presented so far only the top plate is thermostatted and the bottom plate is kept rigid to simplify the application of the constant potential method. Therefore, to confirm the local heating near the electrodes, the effect of the thermostat should be analyzed. Figure \ref{temp_thermo} shows the effect of different thermostat settings on temperature rise and heat removal. For the first setting, where no thermostats are used, the temperature of the system continuously increases. This is because the heat generated in the system is not removed due to the absence of any thermostat and would represent the case of a system between thermally insulating plates. The second setting is where one thermostat is provided at the top plate, and the bottom plate is kept rigid, representing a top plate with high thermal conductivity. Finally, the third setting is used in which both top and bottom plates are thermostatted. Since the bottom plate is usually kept fixed due to theenable efficient application of the CPM, in order to do this an additional layer of flexible gold plate sites is introduced which overlap the the position of the fixed gold plate sites. There are no interactions between the fixed and flexible electrodes sites.  The detailed methodology used for overlapping electrodes is available in our previous work \cite{alosious2024interfacial}. This would represent the case where both plates have high thermal conductivity.  Even though the rate of temperature rise is reduced when a thermostat is applied to the bottom plate, a temperature rise still occurs. From the data showing the heat removed by thermostat data in Fig \ref{temp_thermo}b, it is clear that more heat is removed from the fluid when two thermostats are applied. This is the reason for  a lower rate of temperature rise in the two thermostat systems. 
 
\begin{figure}[h!]
\centering
\includegraphics[width=.6\textwidth]{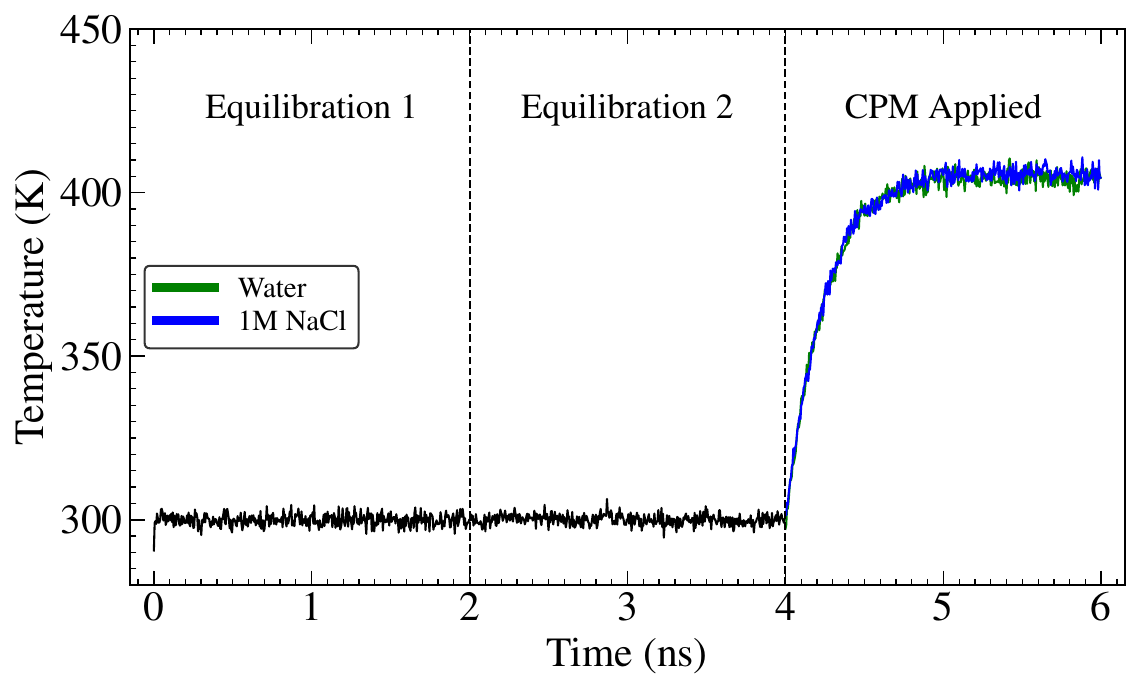}
\caption{Temperature of the fluid as a function of simulation time for water and 1M NaCl solution.  }
\label{temp_water}
\end{figure}

From the above analysis, it is clear that heat is being generated in the fluid nearer to the electrodes when a very high-frequency AC potential is applied. To understand the role of charged particles (Na$^+$ and Cl$^-$ ions) in this phenomenon, the temperature rise of water and aqueous solution of NaCl is compared in Fig \ref{temp_water}. Interestingly, the temperature rise is found to be similar for both systems, with and without NaCl. This suggests that the localized temperature rise is not caused by the movement of ions in the system under rapid changes in electrode polarity. A simple order-of-magnitude diffusion-timescale estimate provides additional physical insight: the characteristic ionic diffusion length during one AC cycle scales as $\ell_d \sim \sqrt{2D/f}$, where $D$ is the ionic diffusivity and $f$ is the AC frequency. For typical ionic diffusivities in water ($D \sim 10^{-9}\,{\mathrm{m}^2\,\mathrm{s}^{-1}}$), ions cannot diffuse across the 7~nm channel within a single cycle above tens of MHz, and at GHz frequencies the per cycle diffusion distance is only on the order of a nanometer. This supports the reduced role of ion-mediated mechanisms at the high frequencies considered here.

To understand the mechanism behind the temperature rise, an orientational order parameter analysis of water molecules has been carried out \cite{de1993physics}. The order parameter is defined as
\begin{equation}
C = \frac{1}{2} \langle 3\cos^2\theta - 1 \rangle
\label{eqnorder}
\end{equation}
where the angle $\theta$ is defined by $\cos\theta = \hat{\textbf{\textit{n}}} \cdot \hat{\textbf{\textit{z}}}$, with $\hat{\textbf{\textit{n}}}$ representing the direction of the water molecule’s dipole moment and $\hat{\textbf{\textit{z}}}$ the unit vector along the z-axis.

Ideally, the reference direction for $\theta$ should align with the applied electric field. However, in our system, the electrodes lie within the same plane (the xy-plane), and the effective field distribution is spatially non-uniform. Therefore, using a fixed axis direction provides a more consistent basis for orientational analysis. Although the applied electric field lies primarily along the x-direction, we analyzed the orientational order parameter with respect to the z-axis ($C_z$) because the water molecules are confined near the xy-plane. The z-direction captures out-of-plane dipole tilting and fluctuations more sensitively, which are critical to revealing the reorientation mechanisms under field perturbations. This approach has been commonly adopted in interfacial water studies to characterize structural disruption and ordering normal to the surface \cite{willard2010instantaneous}.
For this definition, $C = 0$ corresponds to random dipole orientations, $C = 1$ indicates perfect alignment along the z-axis, and $C = -\frac{1}{2}$ represents dipole alignment perpendicular to the z-axis.

\begin{figure}[h!]
\centering
\includegraphics[width=.9\textwidth]{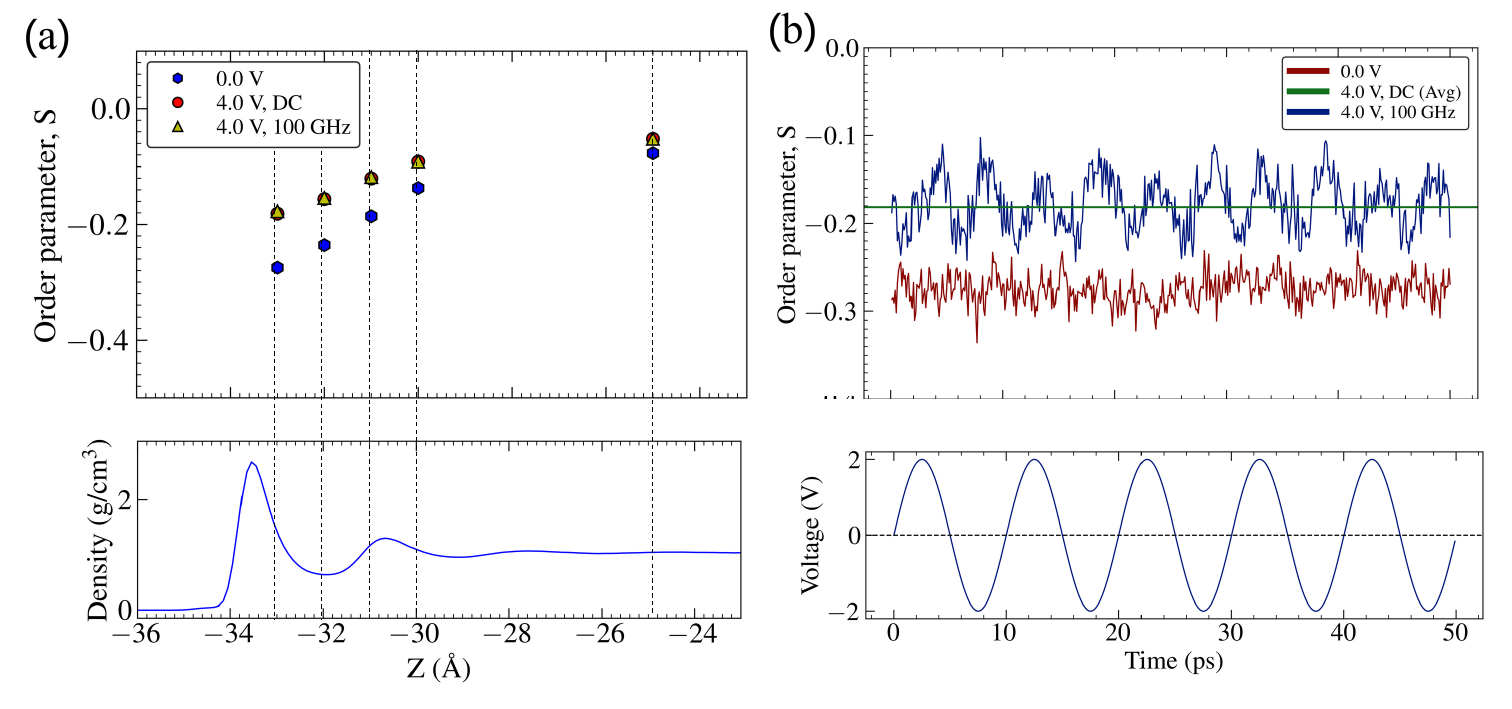}
\caption{Order parameter analysis. (a) The average order parameter for different slabs of fluid near the electrode and the corresponding position of the density profile. Note that the dashed line indicate the boundaries of the slabs, and the order parameter refers to water molecules in the slab to the left of the dashed line.  (b) Instantaneous fluctuation of order parameters for AC and DC potentials along with zero potential. The corresponding voltage variation of AC potential is provided at the bottom. }
\label{orderpara}
\end{figure}

The average orientational order parameter of water dipoles is calculated across different slabs of fluid near the electrode and shown in Fig.~\ref{orderpara}a. The corresponding density profile is also plotted to indicate the interfacial layering near the electrode surface. The highest magnitude of the order parameter is observed in the slab closest to the electrode, indicating strong dipolar alignment near the polarized wall. As the distance from the electrode increases, the order parameter approaches zero, suggesting that the orientational ordering is localized near the interface---consistent with previous studies reporting molecular layering and density enhancement near solid boundaries.
From Fig.~\ref{orderpara}a, it is evident that in the absence of any applied potential, the order parameter with respect to the z-direction ($C_z$) reaches a value of $-0.28$, indicating that water dipoles preferentially align \textit{parallel} to the interface plane (i.e., \textit{perpendicular} to the z-axis). Upon application of an electric potential (either DC or AC), the magnitude of $C_z$ decreases to $-0.10$, reflecting a partial disruption of this in-plane ordering. The dipoles tilt away from their initial configuration, becoming more disordered with respect to the surface normal. This reduction in magnitude suggests that the applied field weakens the surface-induced perpendicular alignment but does not induce a strong net alignment in the z-direction.
To further verify this reorientation behavior, the order parameter with respect to the x-direction ($C_x$) was also computed (not shown in the plot). The analysis showed that $C_x$ decreases from $0.14$ to $0.07$ upon application of the field, supporting the observation that the applied potential leads to an overall loss of orientational order rather than inducing strong alignment along the field direction. For clarity, $C_x$ values are not plotted in the figure.

 Figure \ref{orderpara}b shows the comparison of instantaneous order parameters for AC and DC potentials along with zero potential. In the case of zero potential applied, the instantaneous order parameter randomly fluctuates around -0.28. However, when an AC potential is applied the random fluctuation changes to a periodic fluctuation with twice the frequency of the AC potential. (The reason that the frequency of the order parameter is twice that of the potential frequency will be discussed later.) The average of the magnitude of the order parameter for DC potential is plotted as a straight horizontal line in Fig. \ref{orderpara}b (The fluctuation of DC potential is also random, similar to zero potential). This horizontal line is found to be the baseline for the periodic fluctuation of the order parameter for AC potential. Now, it is clearly understood that when an AC potential is applied, the water molecules will align and re-align, corresponding to the frequency of the AC. Whereas, for DC potential, the alignment of water molecules is inclined towards a particular direction depending upon the polarity.  

\begin{figure}[h!]
\centering
\includegraphics[width=1.0\textwidth]{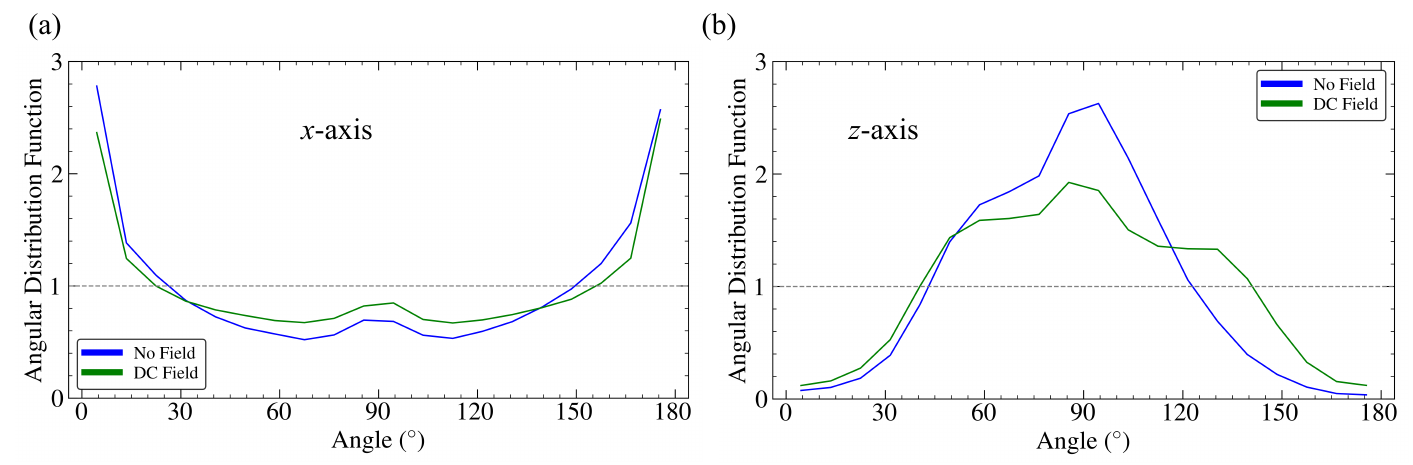}
\caption{The magnitude of the angle between the water dipole moment and the axis direction (a) x-axis (b) z-axis. The angular distribution function was normalized by the surface areas of spherical sections.  }
\label{adf}
\end{figure}

Figure~\ref{adf} shows the angular distribution of water dipole orientations relative to two reference directions: (a) the x-axis and (b) the z-axis. The angular distribution function (ADF) was computed by binning the angle between the molecular dipole vector and the reference axis, and then normalized by the surface area of the corresponding spherical section using a $\sin(\theta)$ correction. This normalization ensures that a uniform (random) dipole orientation yields a flat distribution with $\mathrm{ADF} = 1$, effectively removing geometric bias due to varying surface areas across different angular bins.
In the absence of an applied field, water molecules exhibit a higher likelihood of alignment near $0^\circ$ and $180^\circ$ when measured with respect to the x-axis, as shown in Figure~\ref{adf}a. Conversely, with respect to the z-axis (Figure~\ref{adf}b), dipole orientations are more probable around $90^\circ$, indicating a strong in-plane alignment parallel to the interface. These distributions are consistent with the order parameter values, indicating preferential dipole alignment within the $xy$-plane.
Upon applying a DC electric field, the probability of finding dipoles aligned near $0^\circ$ and $180^\circ$ with respect to the x-axis decreases, suggesting a disruption in the initial dipolar orientation. A similar reduction is observed in the z-direction, indicating a general reorientation of water molecules under the influence of the field. These changes in ADF profiles closely match the observed decrease in order parameter values, thereby confirming that the electric field induces reorientation away from the preferred equilibrium configurations.

\begin{figure}[h!]
\centering
\includegraphics[width=.9\textwidth]{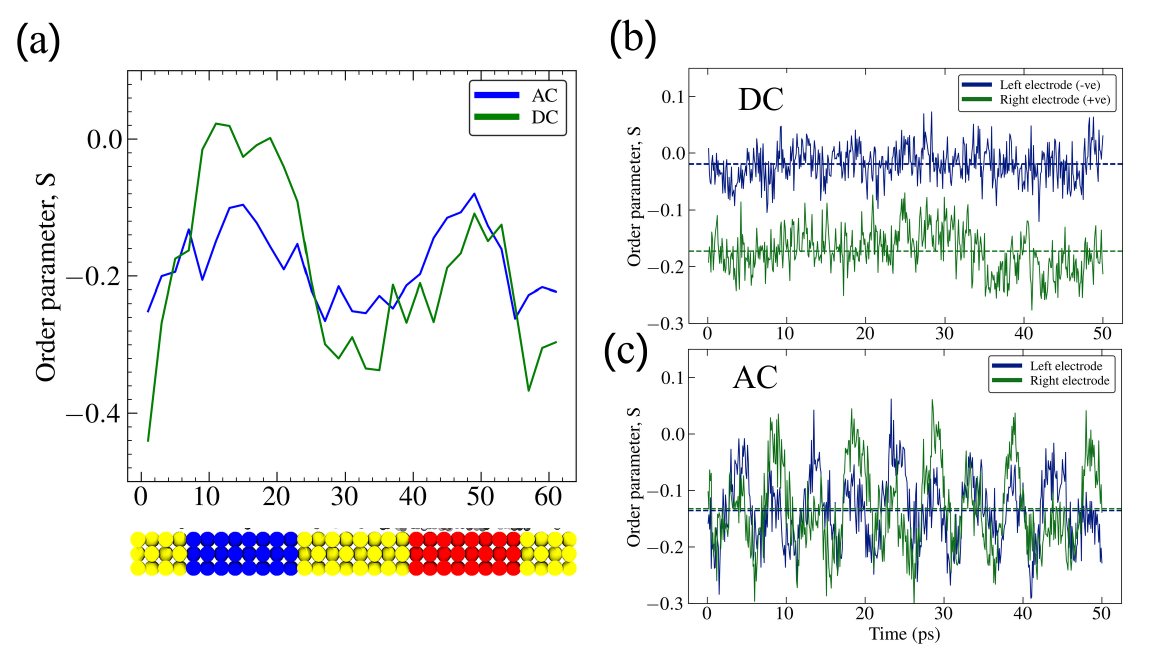}
\caption{The order parameter profile for the first slab of water molecules of the bottom plate with a potential of 4 V. (a) The average value of the order parameter for the slab near the bottom plate. (b) Instantaneous fluctuation of the order parameter on top of left (negative) and right (positive) electrodes under DC potential. (c) Instantaneous fluctuation of the order parameter on top of the left and right electrodes under AC potential.  }
\label{orderpara_electrode}
\end{figure}

To understand this more deeply, the order parameter profile for fluid near the bottom plate for 4 V AC and DC voltages has been analyzed and shown in Fig. \ref{orderpara_electrode}. In this case slabs contain water within 1 {\AA}  of the bottom plate, and in different regions along the plate.  For the DC voltage, the left (blue) and right (red) electrodes correspond to negative and positive charges, respectively. The order parameter profile shows a peak on top of both electrodes, indicating that the maximum deviation from the initial ordering occurs near the electrode positions. Interestingly, the peaks are symmetrical for AC potential and asymmetrical with a large peak on the negative electrode for the DC potential. With this, we can obtain a complete picture of the water molecule ordering near the electrode surface. In the absence of an applied electric field, water molecules at the gold interface exhibit preferential orientation driven by subtle van der Waals and image charge interactions. Studies have shown that the molecular dipole of water tends to lie parallel to the gold surface, resulting in an \textit{in-plane orientation} and a negative order parameter ($C_z < 0$), as also observed in the present study~\cite{limmer2011putative,bonthuis2011dielectric}. 
In this configuration, the oxygen atom (O) typically lies closer to the gold surface, while the hydrogen atoms (H) point slightly away from the surface. This geometry arises because oxygen interacts more favorably with the surface, potentially through weak hydrogen bonding or van der Waals attraction, while the hydrogen atoms avoid direct interaction with gold due to their weaker or slightly repulsive interactions. As a result, the dipole vector (from O to the midpoint of the H--H bond) lies parallel to the surface or is slightly tilted away, leading to a well-defined in-plane ordering in the interfacial water layer.

When an electric potential is applied, the orientation of these dipoles changes in response to the polarity of the electrodes. At the negative electrode, the positively charged hydrogen atoms are attracted towards the surface. This causes the dipoles to tilt in the opposite direction compared to their initial configuration, disrupting the original in-plane alignment. As a result, the order parameter $C_z$ increases significantly towards zero, indicating a strong reorientation or near-random dipole alignment. This explains why, under DC potential, the $C_z$ peak on top of the negative electrode rises to nearly zero.

In contrast, at the positive electrode, the oxygen atoms, which were already slightly tilted towards the surface, experience further attraction due to their negative partial charge. This reinforces the initial orientation rather than disrupting it, resulting in only a modest change in the order parameter. Consequently, the $C_z$ peak on the positive side remains closer to $-0.18$, not reaching zero. This asymmetry in reorientation between the two electrodes under DC voltage accounts for the asymmetric peak heights in the $C_z$ profile observed in Fig.~\ref{orderpara_electrode}. For the AC case, the field reverses direction periodically, which averages out these polarity-specific effects, leading to a more symmetrical response in the order parameter profile.

Finally, the instantaneous order parameter within 1 {\AA} of the electrode plate for the system with an AC potential applied is shown in \ref{orderpara_electrode} b. Earlier it was noted that the frequency of the order parameter is twice that of the potential frequency. The reason for this relationship can be found by analyzing Fig \ref{orderpara_electrode}a and \ref{orderpara_electrode}b. Consider the left electrode; when the polarity of the electrode becomes negative, the maximum peak is achieved, whereas the peak corresponding to positive polarity will be smaller and not significant compared to the average value. Thus, for each cycle of the AC potential, a proportional peak in order parameter is obtained. Similarly, the right electrode also has similar peaks in such a way that the higher peak of the left electrode and the lower peak of the right electrode are in the same phase and vice versa. When the order parameter functions of both electrodes are combined, two peaks are obtained for a single cycle of AC potential. This is the reason why for frequency of the order parameter fluctuation is twice that of the potential frequency. With this detailed order parameter analysis, an in-depth understanding of the behavior of water molecules under AC potential is achieved. Under a very high frequency of AC potential, water molecules will align and re-align due to the periodic change in the polarity of the electrode. This high-frequency periodic ordering of water molecules will create a frictional heat inside the fluid. The water molecules closer to the electrodes will be impacted more and the effect fades as the distance increases. This will create a temperature gradient in the system with a maximum temperature nearer to the electrode as seen in the temperature profile and contour. To ensure that the heating was not due to tuning of the frequency to a resonant frequency, a simulation at 1 THz was carried out and more heating was observed.

From these results we understand that in the system studied here, a very high frequency AC potential will cause work to be done to periodically reorient the water molecules close to the electrode. The energy gained will be dissipated as heat leading to the temperature gradient.  

\begin{figure}[h!]
\centering
\includegraphics[width=1\textwidth]{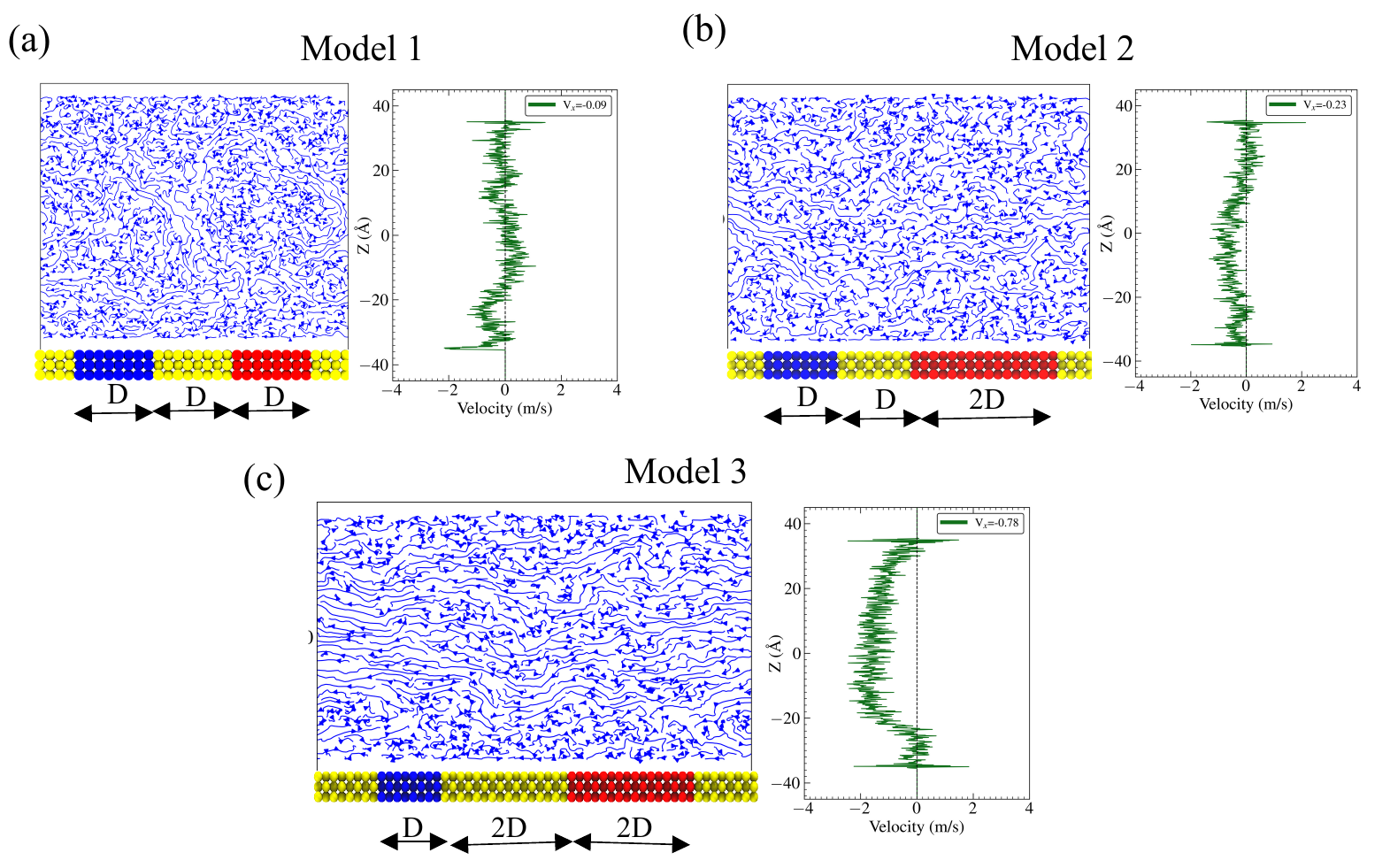}
\caption{Streamline plot and velocity profile of Model 1, Model 2, and Model 3. 
 }
\label{streamline2}
\end{figure}

Until now, we have been focussing on symmetrical electrodes with the same dimensions. It has been already reported that there is a net flow of electrolyte solutions when AC potential is applied to asymmetric pairs of the electrode structure \cite{ramos2003pumping,ajdari1996generation}. Here we focus on three different models to study the influence of asymmetry of electrodes.  The first electrode structure is a symmetric pair of electrodes with both electrodes of the same length D and the distance between the electrodes is also fixed at D. The second structure is an asymmetric electrode in which the left electrode length is D and the right electrode length is 2D (twice the length of left electrode) and the distance between the electrodes is D. Similarly, the third electrode structure is an asymmetric electrode in which the left electrode length is D and the right electrode is 2D, however the distance between electrodes is 2D. These models are named as model 1, model 2, and model 3 respectively. The structure and dimensions of the three models are shown in Fig. \ref{streamline2}. MD simulations were carried out on these three models using the CPM on the electrodes with a 4 V AC potential having a frequency of 100 GHz. The $x$ component of the fluid velocity in the $x$-$y$ plane for the three models is plotted in Fig. \ref{streamline2}. The velocity profile is averaged over five different independent simulations.  Due to the periodicity of the system, even in the cases where the electrodes have different lengths there is symmetry (through reflection + translation).  This means that if flow is produced by the AC voltage, it is equally likely to be in either direction.  Therefore, in averaging the five independent simulations, the average velocities reported are relative to the sign of the overall velocity of the first simulation. 

The average value of the $x$ velocity component in the entire channel is given in the legend of the plots of Fig. \ref{streamline2}. From the velocity profile, it is evident that a net flow in the $x$ direction is obtained for asymmetric electrode configuration. Whereas the net velocity of the symmetric electrode structure is zero,  within the statistical uncertainty. It is also noted that model 3 gives maximum velocity.  The average value of the $x$ component of model 3 is 0.78 $\pm$ 0.34 m s$^{-1}$. A parabolic flow with maximum velocity at the center of the channel is obtained. Similarly, model 2 also shows a net $x$ component velocity of 0.23 $\pm$ m 0.09s$^{-1}$.  

While these velocities are macroscopically significant, the resulting translational kinetic energy per mole of water molecules (5.5 mJ mo$^{-1}$ at speeds of 0.78~m s$^{-1}$) is still six orders of magnitude below the thermal energy scale ($R T=2.5$ kJ mol$^{-1}$ when $T=300$ K and R is the ideal gas constant).

Figure \ref{streamline2} also shows the streamline plot for the three models. The streamlines are more random and chaotic in model 1. The absence of a significant net flow in any direction is also confirmed by the average velocity profile which indicates no net flow. This implies that the forces acting on the system are balanced. It results in a homogeneous distribution of the streamlines with no dominance of the directional movement. 

The streamline plot of Model 2 shows some directional coherence and indicates a net directional flow. This suggests that even though a system might be turbulent, an unbalanced force or gradient may begin to influence the flow direction. The mean velocity profile reveals a slight net flow with a $x$ velocity component of 0.23 m/s. This appears to indicate that the system experiences minor asymmetric forces. 

The third model displays highly directional streamlines, thereby indicating a strong and coherent net flow. This suggests that the forces driving the flow are strong and dominate over turbulent influences. The mean velocity profile shows a net flow in the $x$-direction having a velocity component of 0.78 $\pm$ 0.34 m s$^{-1}$. This net flow in the $x$ direction, together with the streamlined directionality pattern, suggests there is a preferential flow direction. The asymmetries in forces acting on the system drive the fluid preferentially along the $x$ axis over the period considered.

\begin{figure}[h!]
\centering
\includegraphics[width=0.5\textwidth]{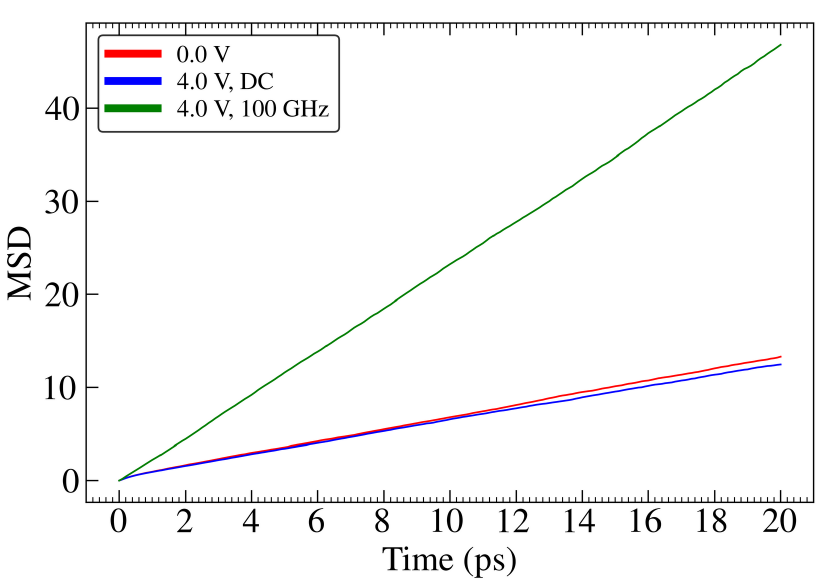}
\caption{The mean square displacement (MSD)  for different field conditions: 0.0 V (no field), 4.0 V DC (direct current field), and 4.0 V at 100 GHz (alternating current field). }
\label{msd}
\end{figure}

The mean square displacement (MSD) of the molecules was calculated to understand the fluid properties, and the results are shown in Fig. \ref{msd}. These systems have been analyzed under different electric field conditions: 0.0 V (no field), 4.0 V DC , and 4.0 V at 100 GHz AC. The MSD plots help to depict how the position of the molecules changes with time. At 0.0 V, a linear increase in MSD indicates normal diffusive behavior. At 4.0 V DC, the MSD is close to that of the no-field condition, indicating minimal impact on the mobility. However, the 100 GHz AC electric field of 4.0 V results in a significantly increased MSD, indicating that the mobility molecules is being enhanced. The increase in MSD observed under 100~GHz AC excitation arises indirectly from the temperature rise induced by the high-frequency electric field. As shown in  \ref{temp_frequency}, no comparable temperature increase or MSD enhancement is observed under 0~V or DC conditions, indicating that the enhanced diffusion is driven by AC-field-induced heating rather than by a direct, non-thermal effect of the electric field on molecular mobility 

Our study aimed to uncover the molecular-level mechanisms responsible for the AC-EHD flow. Several explanations regarding the underlying mechanism of the AC-EHD flow have been proposed in the literature \cite{ramos1998ac, jones1995electromechanics}. Here, by combining the results of our MD simulation and previous understandings, we provide further insight into the mechanism. Various important factors are responsible for this phenomenon. First, localized heat generation occurs near the electrodes due to the effect of the rapid reorientation of water dipoles under the high-frequency AC potential \cite{incropera1996fundamentals}. This localized heating gives rise to a steep temperature gradient between the region near the electrodes and the surrounding cooler fluid. At nanometer length scales, such temperature gradients modify local fluid properties and the response of the fluid to the applied electric field. Under asymmetric conditions, this leads to field-induced fluid motion that enhances transport and mixing
within the channel~\cite{white2011fluid}.

 Secondly, electrothermal effects also play a significant role. The electric field changes the temperature-dependent properties of water, such as its conductivity and permittivity, to give rise to these effects \cite{ramos1998ac}. This gradient in conductivity and permittivity, in turn, interacts with the electric field to produce electrothermal forces that drive the fluid movement. The electrothermal forces increase at higher frequencies, thereby enhancing the fluid movement \cite{castellanos2014electrohydrodynamics}.

In addition to that, Maxwell stress and electrostriction also contribute to the AC-EHD flow. In our simulations, the electric field generated by the asymmetric electrode configurations is non-uniform in both space and time. This results in spatially varying polarization of water molecules across the channel. The order parameter analysis near the electrode surfaces (Figure~\ref{orderpara_electrode}) clearly reveals asymmetrical reorientation behavior under DC and AC fields, with stronger dipole realignment near the negative electrode. This asymmetry implies that the polarization is not uniform, leading to unbalanced dielectric forces characteristic of Maxwell stress.
Moreover, the periodic alignment and realignment of dipoles under high-frequency AC fields, observed in the instantaneous order parameter plots (Figure~\ref{orderpara_electrode}b,c), indicate dynamic electrostriction effects. These rapid dipolar fluctuations effectively deform the local structure of the fluid, especially near the electrodes, resulting in localized stresses. This is further supported by the spatially localized heat generation observed in the temperature contour plots (Figure~\ref{temp_contour}) and the increased mean square displacement (Figure~\ref{msd}) under AC fields. The combination of these localized stresses and temperature-induced softening may contribute additional driving forces to the observed fluid motion. In particular, the emergence of net directional flow in systems with asymmetric electrodes (Figure~\ref{streamline2}) suggests that the imbalance in these Maxwell stress--driven effects is amplified by the geometric asymmetry, further reinforcing their role in AC-EHD flow.

Nevertheless, a net flow is only visible when the electrodes are asymmetric.  In symmetric electrode configurations, the aforementioned forces are balanced, thereby eliminating the chances of a net directional flow.  The difference in the electrode dimensions and the distance between electrodes will enhance the imbalance in forces acting on the fluid, leading to a pronounced AC-EHD effect. Thus, the high-frequency AC potential enhances molecular mobility while inducing localized heating near the electrodes. The combined influence of field-induced thermal effects and non-uniform electric fields drives fluid motion under AC-EHD conditions. Notably, asymmetry in the electrode configuration is essential for creating an imbalance in these effects, resulting in a net directional flow.
The present results provide qualitative guidance for the design of nanoscale AC-EHD
devices. The simulations indicate that operation at very high AC frequencies is essential to activate ion-independent flow under strong confinement, while geometric asymmetry in electrode configurations is a necessary condition for generating directed motion. In addition, the weak dependence of the flow on ionic concentration suggests potential advantages for applications involving complex or variable electrolytes, such as biosensing and lab-on-a-chip systems. The highly localized nature of the flow and heating near electrode surfaces further suggests opportunities for targeted transport or mixing close to functional interfaces rather than bulk stirring.

\section{CONCLUSIONS}

In summary, we have investigated the molecular-level mechanism of AC-EHD flow in a gold-NaCl nanoscale system under very high AC frequencies using classical molecular dynamics simulations. An aqueous NaCl solution was confined between two gold nanoplates, with the bottom plate divided into two electrodes to which a 4.0 V AC potential was applied using the constant potential method. Frequencies up to 100 GHz were explored to ensure complete field cycles within the nanosecond simulation timescale.

A significant temperature rise in the fluid was observed above 10 GHz, leading to a steep temperature gradient from the bottom to the top plate. Notably, this temperature increase was found to be largely independent of the presence of Na$^+$ and Cl$^-$ ions, indicating that the primary mechanism of heat generation arises from the water molecules themselves. Through order parameter and angular distribution analyses, we showed that water dipoles periodically align and re-align in response to the oscillating field. This cyclic reorientation under high frequency creates frictional heating, especially near the electrodes, resulting in localized thermal gradients.

To evaluate fluid motion, three different electrode configurations were studied. The simulations revealed that asymmetric electrode structures produced a net directional flow, while symmetric configurations did not. The strongest flow was observed in the most asymmetric case, with a parabolic velocity profile and peak velocity of 0.78 m s$^{-1}$. Streamline plots and MSD analyses further confirmed enhanced particle mobility and flow coherence under AC fields in asymmetric systems.

Several physical mechanisms were identified as contributors to this flow: (1) field-induced effects associated with temperature-induced density gradients, (2) electrothermal forces from the interaction between the electric field and temperature-dependent properties of water, and (3) Maxwell stress2) stresses arising from non-uniform electric fields and electrostriction caused by spatially varying polarization. The observed flow is enabledemerges only when these forceseffects are imbalanced, which occurs under geometrically asymmetric electrode configurations.

These results also suggest that the dynamic reorientation of water dipoles plays a central role in driving AC-EHD flow under the conditions examined. However, such reorientation alone is not sufficient to produce a net directional flow unless it occurs collectively. If the dipoles reorient randomly, some tilting to the left and others to the right, the resulting forces would cancel out, preventing sustained movement. On the other hand, if the water molecules tend to tilt in the same direction (e.g., from a perpendicular to a $30^\circ$ alignment), the collective motion can reinforce local fluid deformation and contribute to net flow. This coordination may arise from the coupling between neighboring dipoles and is likely sensitive to initial molecular orientations. Under high-frequency AC fields, this initial preference may persist across cycles, preventing the system from relaxing back to its equilibrium configuration and resulting in a time-averaged directional force. This provides a plausible molecular-level explanation for why net flow is observed only under specific electrode asymmetries and high-frequency conditions in our simulations.

These insights offer a comprehensive picture of AC-EHD flow at the nanoscale and help to uncover the underlying molecular mechanism. They are crucial for optimizing electrode design and field parameters in nanofluidic systems. In future work, a quantitative analysis of the above-mentioned forces using a polarizable water model will be carried out to gain a deeper understanding and predictive capability. 

Finally, we note that the observations here are for a nanoscale system (both the spacing between the electrodes and the pore width are just nanometers) and very high frequency AC fields. As frequency is reduced and/or the system size increases we expect other factors to play a role and perhaps become more dominant.  For example, heat generation might vary with position on the electrode as the electrode becomes larger, the ions could have time to  reorder or flow if the frequency is reduced, nanoscale features observed here might not be resolved when measured in a larger system and convective or turbulent flow could be observed at lower driving forces. How different factors contribute as the scales change provides interesting challenges for future work.

\begin{acknowledgments}
This research was supported by the Australian Research Council
through the ARC Laureate Fellowship program (FL190100080 and FL220100059). It was also supported by resources provided by the Pawsey Supercomputing Research Centre with funding from the Australian Government and the Government of Western Australia and resources and services from the National Computational Infrastructure (NCI) under the Adapter scheme, which is supported by the Australian Government. We also wish to acknowledge The University of Queensland’s Research Computing Centre (RCC) and Queensland Cyber Infrastructure Foundation (QCIF) for their support.
\end{acknowledgments}

\bibliography{reference}% Produces the bibliography via BibTeX.

\end{document}